\documentclass[fleqn,10pt]{wlscirep}
\usepackage[utf8]{inputenc}
\usepackage[T1]{fontenc}
\usepackage{physics}
\usepackage{float}
\usepackage{qcircuit}
\usepackage{braket}
\usepackage{amsmath}
\usepackage{amsfonts}
\usepackage{multirow}
\usepackage{graphicx}
\usepackage{subcaption} 
\usepackage{pdfpages}
\usepackage{pgffor}
\usepackage{tikz}
\definecolor{myblue}{HTML}{4f6fd6}
\usetikzlibrary{quotes,shapes.geometric,arrows,fit,positioning,arrows.meta}
\tikzstyle{action} = [rectangle, rounded corners, 
minimum width=3cm, 
minimum height=1cm,
text centered, 
draw=black, 
inner sep=2mm,
text width=4cm,
fill=myblue,
text=white]

\tikzstyle{container} = [rectangle,
rounded corners, 
minimum width=3cm, 
minimum height=1cm, 
text centered, 
text width=3cm, 
draw=myblue,
thick]

\tikzstyle{label} = [rectangle,
rounded corners, 
minimum width=3cm, 
minimum height=1cm, 
text centered, 
draw=myblue,
text=myblue,
thick,
fill=lightgray!50]
\tikzstyle{arrow} = [thick,-{Latex[length=4mm]}]

\title{Solving nonlinear differential equations on noisy $156$-qubit quantum computers}

\author[1,a]{Karla Baumann}
\author[1,b]{Youcef Mohdeb}
\author[1,b]{Roman Randrianarisoa}
\author[1,b]{Roland Katz}
\author[1,b]{Aoife Boyle}
\author[1,2,3,c]{Frédéric Holweck}
\affil[1]{ColibriTD, Paris, 75000, FRANCE}
\affil[2]{Laboratoire Interdisciplinaire Carnot de Bourgogne,  UMR 6303 CNRS, University of Technology of Belfort-Montb\'eliard, 90010 Belfort Cedex, France}
\affil[3]{Mathematics and Statistics Department, Auburn University, Auburn, AL, USA}
\affil[a]{karla.baumann@colibritd.com (corresponding author)}
\affil[b]{quantum@colibritd.com}
\affil[c]{frederic.holweck@utbm.fr}

\begin{abstract}
In this paper, we report on the resolution of nonlinear differential equations using IBM’s quantum platform. More specifically, we demonstrate that the hybrid classical–quantum algorithm H-DES successfully solves a one-dimensional material deformation problem and the inviscid Burgers' equation on IBM’s 156-qubit quantum computers. These results constitute a step toward performing physically relevant simulations on present-day Noisy Intermediate-Scale Quantum (NISQ) devices. 
\end{abstract}
\begin{document}

\flushbottom
\maketitle

\thispagestyle{empty}

\section*{Introduction}

The numerical solution of partial differential equations (PDEs) forms a cornerstone of modern scientific computing, underpinning simulation and modeling across a wide spectrum of scientific and industrial applications. It is therefore unsurprising that, in the early days of digital computing, the Los Alamos team, under the leadership of John von Neumann \cite{aspray1990john}, undertook some of the first numerical experiments on the ENIAC computer to solve simple ordinary and partial differential equations. The remarkable diversity of numerical methods and advanced computational tools available today stands as a testament to the vision and pioneering efforts of these early researchers.

We are now entering the early stages of the quantum computing era, a new paradigm of digital computation that exploits the fundamental properties of subatomic particles  to perform calculations in ways fundamentally different from classical computers. Although quantum algorithms have been theoretically known since the 1980s\cite{deutsch1985quantum}, their practical implementation on real quantum hardware has only recently become feasible with the advent of Noisy Intermediate-Scale Quantum (NISQ) devices. These emerging systems, accessible through cloud-based platforms such as IBM Quantum and Amazon Braket mark an important step toward realizing the potential of quantum computation. Just as researchers in the early era of classical computing asked whether differential equations could be solved numerically, one may now ask: can we solve ordinary and partial differential equations on quantum computers? 

The possibility of accelerating classical numerical methods, such as the Finite Element Method (FEM), using quantum computation was established at the theoretical level by Montanaro and Pallister \cite{montanaro2016quantum}. Their approach relies on the Harrow–Hassidim–Lloyd  quantum algorithm for solving systems of linear equations, a conceptually elegant result that, however, remains far beyond the reach of current  hardware.  

This limitation has motivated the use of Variational Quantum Algorithms (VQAs) for solving ordinary and partial differential equations, explored in several recent works and under various formulations \cite{kyriienko2021solving,sarma2024quantum,hunout2025variational,donachie2025solving}. However, these studies have so far demonstrated their proof-of-concept implementations primarily on  quantum simulators.

In this paper, we report the first successful resolution of non-trivial differential equations on an actual quantum hardware platform, namely the IBM Quantum processors. Our approach is based on the implementation of the H-DES algorithm\cite{jaffali2024h}, the principles of which will be detailed in the Method section.

The first differential problem we successfully solved on IBM Quantum hardware is the one-dimensional hypoelastic tensile test for material deformation. In this benchmark 
problem, the material is modeled as a one-dimensional bar, fixed at one end and subjected 
to a tensile load at the other. The one-dimensional hypoelastic tensile test is a standard problem in solid mechanics, involving a 
coupled system for displacement and stress, a nonlinear constitutive contribution with 
several material parameters, and boundary conditions corresponding to a realistic loading 
scenario. The governing system of ordinary differential equations 
for this classical setting is

\begin{equation}\label{eq:1dmaterial}
\left\{
\begin{array}{l}
u'(x) - \dfrac{\sigma(x)}{K} 
    - \dfrac{2}{\sqrt{3}}\,\epsilon_0 
      \left( \dfrac{\sigma(x)}{\sqrt{3}\,\sigma_0} \right)^{n} = 0, \\[8pt]
\sigma'(x) + b = 0 ,
\end{array}
\right.
\end{equation}

where \(u(x)\) denotes the displacement field and \(\sigma(x)\) the stress field with $u'$ and $\sigma'$ the corresponding first order derivatives.  
The parameter \(K\) represents the bulk modulus of the material, \(n\) is the exponent 
of the power-law term, \(b\) is the applied force per unit mass, \(\epsilon_0\) is the 
proportional strain limit, and \(\sigma_0\) is the proportional stress limit.
Although the system \eqref{eq:1dmaterial} admits a closed-form analytical solution and 
is therefore relatively simple from the standpoint of classical numerical analysis, it 
constitutes a nontrivial and physically meaningful benchmark for a first demonstration 
of a quantum differential equation solver on real hardware.  The combination of analytical tractability and physical relevance provides an 
ideal test case: it allows for a precise comparison between the quantum solution and the 
exact reference, while still capturing key structural features (nonlinearity, parameter 
dependence, and coupling of fields) that appear in more complex PDEs and FEM-based 
models of material deformation.

The second differential problem implemented on real quantum hardware in this work, 
is the inviscid Burgers' equation. This partial differential equation models the 
evolution of an isobaric, incompressible and non-viscous fluid flow and is given by
\begin{equation}\label{eq:inviscid}
\frac{\partial u(x,t)}{\partial t} + 
u(x,t)\,\frac{\partial u(x,t)}{\partial x} = 0,
\end{equation}
where \(u(x,t)\) denotes the fluid velocity field. To obtain a well-posed problem, an initial condition, $u(x,0)$, must be prescribed.
The inviscid Burgers' equation is challenging for different reasons than the hypoelastic ODE. The difficulty arises from the increase in dimensionality  and the nonlinear advection term 
\(u\,\partial_x u\), which induces the formation of steep gradients and, eventually, shock waves 
in finite time, even when starting from smooth initial conditions. The inviscid Burgers' equation is an example of a first-order quasi-linear hyperbolic equation that has long served as a 
canonical test case for numerical schemes aiming to capture nonlinear hyperbolic 
dynamics.

\section*{Method: H-DES, a toolbox for solving PDEs  on quantum computers}

Today, in the era of NISQ computers, the most 
promising pathway to obtaining practical results on real quantum hardware lies in the 
use of VQAs. A VQA consists of a parameterized quantum 
circuit designed to prepare a quantum state \(\ket{\psi(\theta)}\), where the vector of 
parameters \(\theta\) controls the action of the circuit. These parameters are optimized 
classically through an iterative procedure driven by a loss function that encodes the 
constraints or conditions the quantum state must satisfy in order to solve the target 
problem. In this hybrid quantum--classical paradigm, the quantum device prepares the 
states and evaluates expectation values, while the classical optimizer adjusts the 
parameters, allowing VQAs to operate within the hardware limitations characteristic of 
current NISQ systems.

\begin{figure}[!ht]
\begin{center}
  \includegraphics[width=0.85\linewidth]{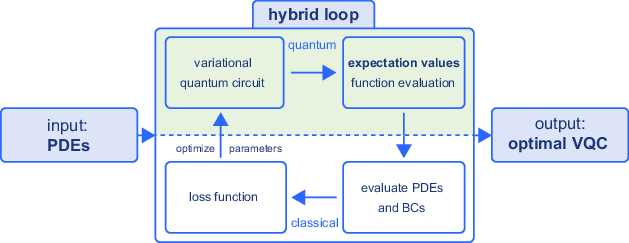}
\caption{H-DES workflow and hybrid loop: The input PDEs and BCs  define the loss function that guides  the optimization  process. The solution is encoded in the VQC. The loop stops when the algorithm reaches an acceptable value of the loss function which indicates a potentially  acceptable approximate solution of the problem.}\label{fig:hdes}
\end{center}
\end{figure}

The H-DES algorithm \cite{jaffali2024h}, which we employ to solve both the 1-D material 
deformation problem and the inviscid Burgers' equation, is a hybrid classical--quantum 
differential equation solver that leverages the expressivity of variational quantum 
circuits (VQCs) to encode solutions of ODEs and PDEs. The algorithm takes as input a 
differential equation together with its associated boundary or initial conditions. These 
elements define a loss function that guides the hybrid optimization loop. Within this 
loop, the VQC prepares a quantum state \(\ket{\psi(\theta)}\) intended to represent the 
solution of the differential problem. After measurement, the resulting expectation 
values are used to evaluate the loss, and the circuit parameters \(\theta\) are updated 
via a classical optimization step. The process iterates until the loss falls below a prescribed tolerance, indicating convergence toward an approximate solution of the 
original differential equation. The global workflow of H-DES is summed up in Figure \ref{fig:hdes}.

\subsection*{Variational ansatz, evaluation and encoding of the solution}

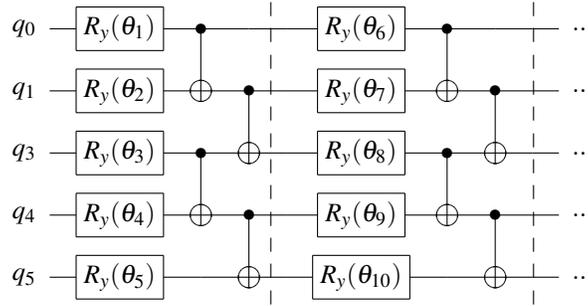
\begin{figure}[!ht]

\centerline{
\Qcircuit @C=1em @R=.7em {
  \lstick{q_0} & \gate{R_y(\theta_1)} & \ctrl{1} & \qw      & \qw \barrier[-1.2cm]{4} & \gate{R_y(\theta_6)}  & \ctrl{1} & \qw      & \qw \barrier{4} & \rstick{...} \qw \\
  \lstick{q_1} & \gate{R_y(\theta_2)} & \targ    & \ctrl{1} & \qw                     & \gate{R_y(\theta_7)}  & \targ    & \ctrl{1} & \qw             & \rstick{...} \qw \\
  \lstick{q_3} & \gate{R_y(\theta_3)} & \ctrl{1} & \targ    & \qw                     & \gate{R_y(\theta_8)}  & \ctrl{1} & \targ    & \qw             & \rstick{...} \qw \\
  \lstick{q_4} & \gate{R_y(\theta_4)} & \targ    & \ctrl{1} & \qw                     & \gate{R_y(\theta_9)}  & \targ    & \ctrl{1} & \qw             & \rstick{...} \qw \\
  \lstick{q_5} & \gate{R_y(\theta_5)} & \qw      & \targ    & \qw                     & \gate{R_y(\theta_{10})} & \qw      & \targ    & \qw             & \rstick{...} \qw 
}}
\caption{Hardware-efficient ansatz used in H-DES\cite{jaffali2024h}.}
\label{fig:ansatz}
\end{figure}

The VQC (or ansatz) of H-DES  transforms the input state \(\ket{0}^{\otimes n}\) into the 
parameterized state, written in the computational basis,
\[
\ket{\psi(\theta)}=\sum_{i \in \{0,1\}^n} a_i(\theta)\ket{i},
\]
where the amplitudes \(a_i(\theta)\) depend on the parameters \(\theta\) defining the 
rotation gates within the circuit. The standard ansatz used in H-DES is a simplified version of the so-called Hardware-Efficient ansatz\cite{jaffali2024h}. Our hardware-efficient structure is composed of layers of \(R_Y\) rotation gates followed by CNOT gates (see Figure~\ref{fig:ansatz}), 
arranged according to the connectivity constraints of the underlying quantum processor. 

The choice of ansatz may be adapted to the specific differential problem being addressed. 

To encode the solution functions $f_j$ of the PDE and their derivatives, $f_j^{(m)}$, we define a VQC for each value $j$ and denote $\ket{\psi_j(\theta)}$, the resulting quantum state. We also consider an observable $\mathcal{O}_{m}(x)$ such that the function values $f^{(m)}_j(x)$ are the expectation values of this observable at $\ket{\psi_j({\theta})}$.
\begin{equation}
f_j^{(m)}(x) = \lambda_j\langle \psi_j(\theta)|\mathcal{O}_{m}(x)| \psi_j(\theta)\rangle.
\end{equation}
 Denoting by \(\lambda_j\) a scaling parameter and $Z$ the usual $Z$-Pauli matrix, we  can define $\mathcal{O}_{m}$ as
\begin{equation}\label{eq:globalobs}
\mathcal{O}_{m}(x) = Z \otimes 
\left(
\sum_{i=0}^{2^n-1} \mathrm{Cheb}^{(m)}(i,x)\, \ket{i}\!\bra{i}
\right).
\end{equation}
This choice of observable corresponds to a Chebyshev expansion of the function $f_j$, where $\mathrm{Cheb}(i,x)=\cos(n\arccos{x})$ are the usual Chebyshev polynomials.

It is important to note that H-DES differs fundamentally from the VQC methods proposed in other works\cite{kyriienko2021solving,hunout2025variational}, where the encoding of collocation 
points is achieved through a feature map applied directly within the quantum circuit. 
By contrast, postponing the spectral decomposition to the observable offers additional 
flexibility: the observables \(\mathcal{O}_m\) themselves may be varied or optimized. Another possible choice of observables than Eq.~(\ref{eq:globalobs}) is to use linear combination of local Pauli monomials. A $k$-local Pauli monomial is an observable $P^k$ defined on $n$-qubit such that 
\(P^k=P_1\otimes\dots\otimes P_n\in \{X,Y,Z,I\}^{\otimes n}\) and where a minimum of  $n-k$ atoms $P_j$ are identity operators. Considering $P^{\: k,i}$ a chosen family of $k$-local Pauli monomials, one may define the $k$-local observable $\mathcal{O}_{C}(x)$ 
\begin{equation}
\mathcal{O}_C(x) = \sum_{i=1}^M \alpha_i\, \mathrm{Cheb}(i,x)\, P^{k,i}, \text{ with }\alpha_i \in \mathbb{R}, \text{ and } C=\{P^{k,i},\alpha_i\}.
\end{equation}

The use of local observables makes our VQC less prone to barren plateaus\cite{cerezo2021cost, jaffali2024h}.
Note that, as we avoid the use of feature maps, the evaluation of the derivatives of $f_j$ is done with the same circuit.\\

The encoding of multivariate functions for PDEs follows the same strategy by defining as many registers as variables\cite{jaffali2024h}.

\subsection*{Loss construction and boundary-condition encoding in H-DES}

Once the evaluation of the function is encoded in the choice of our VQC, one needs to build the loss function, which 
quantifies how well the function evaluated by the VQC satisfies the target differential
equation together with its associated boundary or initial conditions. For a generic  differential operator \(\mathcal{D}\) and unknown function \(f_\theta(x)\), the residual
\begin{equation}
R(x,\theta) = \mathcal{D}f_{\theta}(x)
\end{equation}
is used to define a loss term of the form
\begin{equation}
\mathcal{L}_{\mathrm{PDE}}(\theta) =
\frac{1}{|S|}\sum_{x\in S} \big| R(x,\theta) \big|^2, \text{ where $S$ denotes a set of collocation points.} \
\end{equation}

This term enforces fidelity 
to the differential equation, while additional contributions must ensure that the 
solution satisfies the required boundary conditions.

Two mechanisms allow to incorporate boundary conditions. The 
first is to encode them directly within the loss function, for example by adding penalty 
terms of the form
\begin{equation}
\mathcal{L}_{\mathrm{BC}}(\theta) = 
\sum_{k} \big| f_{\theta}(x_k) - g_k \big|^2,
\end{equation}
where \(g_k\) denotes the prescribed boundary values. 

The second mechanism consists of a 
floating boundary condition shift, in which the trial function is modified so that the 
boundary conditions are automatically satisfied by construction. For instance, one may 
represent the solution as a shifted function 
\(f_{\theta}(x) = h(x) + s(x)\), where \(s(x)\) enforces the boundary constraints and 
the variational ansatz is only responsible for approximating the homogeneous component 
\(h(x)\).

 Once the full loss 
function is assembled,
\begin{equation}
\mathcal{L}(\theta) = \mathcal{L}_{\mathrm{PDE}}(\theta) 
+ \lambda_{\mathrm{BC}} \mathcal{L}_{\mathrm{BC}}(\theta) 
\end{equation}

with $\lambda_{BC}$ a weight to balance the effect of the boundary conditions in the loss function, the parameters \(\theta\) are optimized  using standard classical optimizers such as 
Adam\cite{KingmaBa2015Adam} and L-BFGS\cite{LiuNocedal1989LBFGS} or  gradient-free methods like COBYLA\cite{Powell1994COBYLA} and CMA-ES\cite{HansenOstermeier2001CMAES} common in variational quantum 
algorithms. 

The variety of possible choices in running H-DES (including the structure of the 
ansatz, the selection of observables and basis functions for encoding or evaluating the solution, and the 
strategy used to impose boundary conditions)  highlights the toolbox character of the 
method: depending on the structure of the  problem, one may flexibly explore 
different modeling and optimization options.

\subsection*{Quantum hardware--specific settings}

When executing variational quantum algorithms on real hardware, a central design choice 
concerns the architecture of the parameterized quantum circuits. In this work, we employ 
a hardware-efficient ansatz (HEA), Fig. \ref{fig:ansatz}, and investigate several circuit architectures based on 
parameterized single-qubit rotation gates combined with two-qubit entangling gates, 
including CNOT, CZ, and ECR gates. The design of these circuits must be considered in the 
context of the full real-hardware execution workflow, taking into account effects on 
execution time, transpilation overhead, and the potential use of noise suppression or 
error mitigation techniques.

All experiments were performed on the \texttt{ibm\_marrakesh} and \texttt{ibm\_fez}
devices, which belong to the 
latest generation\footnote{At the time of the experiments: May and December 2025.} of IBM Quantum processors based on the Heron architecture. To run the designed VQC on real backend, the circuit goes through a transpilation process which adapts the algorithm to the topology (qubits connectivity) of the real machine. 
The IBM Quantum Platform\footnote{https://quantum.cloud.ibm.com/} proposes different options of transpilation where a level of optimization (for the number of gates and depth of the circuit) and qubit allocation (to map the qubits  of the circuit to the physical qubits  of the machine) can be chosen. For circuit transpilation, we used the Qiskit compilation pipeline  generated by the command
\href{https://docs.quantum.ibm.com/api/qiskit/transpiler_preset}{\texttt{generate\_preset\_pass\_manager()}}\footnote{https://docs.quantum.ibm.com/api/qiskit/transpiler\_preset}. To keep the running time low, we choose a medium optimization of the circuit and a trivial allocation of the qubits with the commands \href{https://quantum.cloud.ibm.com/docs/en/guides/set-optimization}{\texttt{optimization\_level = 2}}\footnote{{https://quantum.cloud.ibm.com/docs/en/guides/set-optimization}} and \href{https://quantum.cloud.ibm.com/docs/en/guides/transpiler-stages}{\texttt{layout\_method = ``trivial''}}\footnote{{https://quantum.cloud.ibm.com/docs/en/guides/transpiler-stages}}.  Several 
alternative transpilation strategies were tested, but this configuration proved to be 
the most robust for our circuits and for the requirements of our 
workflow.  Due to the stochastic nature of the transpilation process when numerical values are involved in gate parameters, this step was instead performed only once during the preprocessing phase using symbolic values for gate parameters. The resulting transpiled circuit, 
with symbolic parameters retained, was then stored in memory and reused throughout the 
entire optimization loop in order to ensure a stable and reproducible training process.

To minimize 
gate noise, it is possible to configure the error mitigation module of the IBM Quantum platform by choosing a resilience level for the measurement estimator. Higher level of resilience provides more accurate results but implies a longer processing time. In our computation the convergence of the 
variational optimization was achieved without the need for advanced noise suppression or 
error mitigation strategies. All runs, presented here, relied  on 
\href{https://docs.quantum.ibm.com/guides/configure-error-mitigation}{\texttt{resilience\_level = 0}}\footnote{https://docs.quantum.ibm.com/guides/configure-error-mitigation}.

\section*{Results}

In this section we report real-hardware results obtained with H-DES on IBM Quantum 
processors for two benchmark problems described in the Introduction: (i) a nonlinear hypoelastic system of ODEs 
arising in 1-D material deformation, and (ii) the inviscid Burgers' equation. While we 
adopt a common experimental philosophy across both benchmarks, the concrete choices of 
ansatz, optimizer, and noise-handling strategy differ, reflecting the distinct numerical 
structure and hardware sensitivity of the underlying differential problems.  For each 
benchmark, we present the differential formulation and reference solution, the H-DES configuration, and the convergence and accuracy obtained on 
hardware.

\subsection*{Nonlinear hypoelastic 1-D tensile test\\}
\vspace{-2mm}
\paragraph{Model and analytical reference.}
We consider the nonlinear system of ODEs
\begin{equation}\label{eq:md_system}
\left\{
\begin{array}{l}
u' - \dfrac{\sigma}{K} 
    - \dfrac{2}{\sqrt{3}}\,\epsilon_0 
      \left( \dfrac{\sigma}{\sqrt{3}\,\sigma_0} \right)^{n} = 0, \\[8pt]
\sigma' + b = 0 ,
\end{array}
\right.
\end{equation}
with boundary conditions \(u(0)=0\) and \(\sigma(0)=g\) (the stress at the origin). This system admits a closed-form
analytical solution, $u$ is a degree $n+1$ polynomial and $\sigma$ is a degree $1$ polynomial, which we use as a ground-truth reference for validation.

\paragraph{Ansatz selection (real-hardware constraints).}
For this benchmark we used the HEA architecture 
combining parameterized single-qubit rotations with  two-qubit CNOT gates (Fig.~\ref{fig:ansatz}). 
Because the loss involves both the solution functions and their derivatives, sufficient 
expressivity must be achieved without excessive circuit depth. Empirically, shallow 
layered HEAs provided the best trade-off between expressivity, noise accumulation, and 
optimization stability. For the results reported here, we use a layered 
HEA with depths 2 and 4 for the $15$ qubits circuit
representing \(u(x)\) and the $4$ qubits circuit representing \(\sigma(x)\), respectively. 

\paragraph{Boundary-condition strategy and loss.}
We enforce boundary conditions exactly through a floating boundary-condition shift:
\begin{equation}
\hat{u}_{\theta}(x) = u_{\theta}(x) - u_{\theta}(0), 
\qquad 
\hat{\sigma}_{\theta}(x) = \sigma_{\theta}(x) - \sigma_{\theta}(0).
\end{equation}

Let \(\mathcal{D}_1\) and \(\mathcal{D}_2\) denote the differential operators associated with the 
first and second equations in \eqref{eq:md_system}. We minimize the summed squared residuals
\begin{equation}\label{eq:md_loss}
\mathcal{L}_{\text{PDE}}(\theta)=\dfrac{1}{|S|}\sum_{x\in S}\Big(
\mathcal{D}_1(\hat{u}_{\theta}(x), \hat{\sigma}_{\theta}(x))^2
+
\mathcal{D}_2(\hat{\sigma}_{\theta}(x))^2
\Big),
\end{equation}

where \(S\) denotes a set of collocation points.

\paragraph{Optimizer and gradients.}

We employ SLSQP a gradient-based optimizer. Gradients are evaluated on hardware using the parameter-shift rule \cite{wierichs2022general}. To evaluate the gradient of such a cost function, we note that
\begin{equation} 
\label{Gradient}
\partial_{\theta} \mathcal{L}_{\text{PDE}}(\theta)= 2 \sum_{x\in S} (\partial_{\theta} \mathcal{D}_1) \mathcal{D}_1(\hat{u}_{\theta}(x),  \hat{\sigma}_{\theta}(x))+  (\partial_{\theta} \mathcal{D}_2) \mathcal{D}_2(\hat{\sigma}_{\theta}(x)).
\end{equation}
 
 For each circuit parameter \(\theta_j\), the derivative of any 
measured expectation value is obtained via
\[
\partial_{\theta_j}\langle \mathcal{O}\rangle
=
\frac{\langle \mathcal{O}\rangle_{\theta_j+\pi/2}-\langle \mathcal{O}\rangle_{\theta_j-\pi/2}}{2},
\]
where $\langle \mathcal{O}\rangle$ is the expectation value of the chosen observable to represent $\hat{u}_{\theta}(x)$.  The gradient of the total loss is obtained by the chain rule by first reconstructing \(\partial_{\theta_j}\mathcal{D}_i\) from $\partial_{\theta_j}\langle \mathcal{O}\rangle$. Note that this requires \(2\times\) (number of circuit parameters) 
additional circuit evaluations per iteration. 

\paragraph{Observable encoding.}
To illustrate the flexibility of H-DES, we use a mixed encoding: a 1-local Pauli observable for \(u(x)\),
\begin{equation}
\mathcal{O}_C(x)=\sum_{i=0}^{n_1-1}\mathrm{Cheb}(i,x)\,Z_i, \text{ (where } Z_i \text{ is a Pauli monomial with  $Z$ on qubit $i$ and the identity everywhere else)}
\end{equation}
and a global observable for \(\sigma(x)\),
\begin{equation}
\mathcal{O}(x)=
Z\otimes \left(\sum_{i=0}^{2^{n_2}-1}\mathrm{Cheb}(i,x)\ket{i}\!\bra{i}\right),
\end{equation}
implemented via two separate circuits using \(n_1\) and \(n_2\) qubits respectively.

\begin{figure}[h!]
\centering
\begin{minipage}[t]{0.333\linewidth}
  \centering
  \includegraphics[page=1,width=\linewidth]{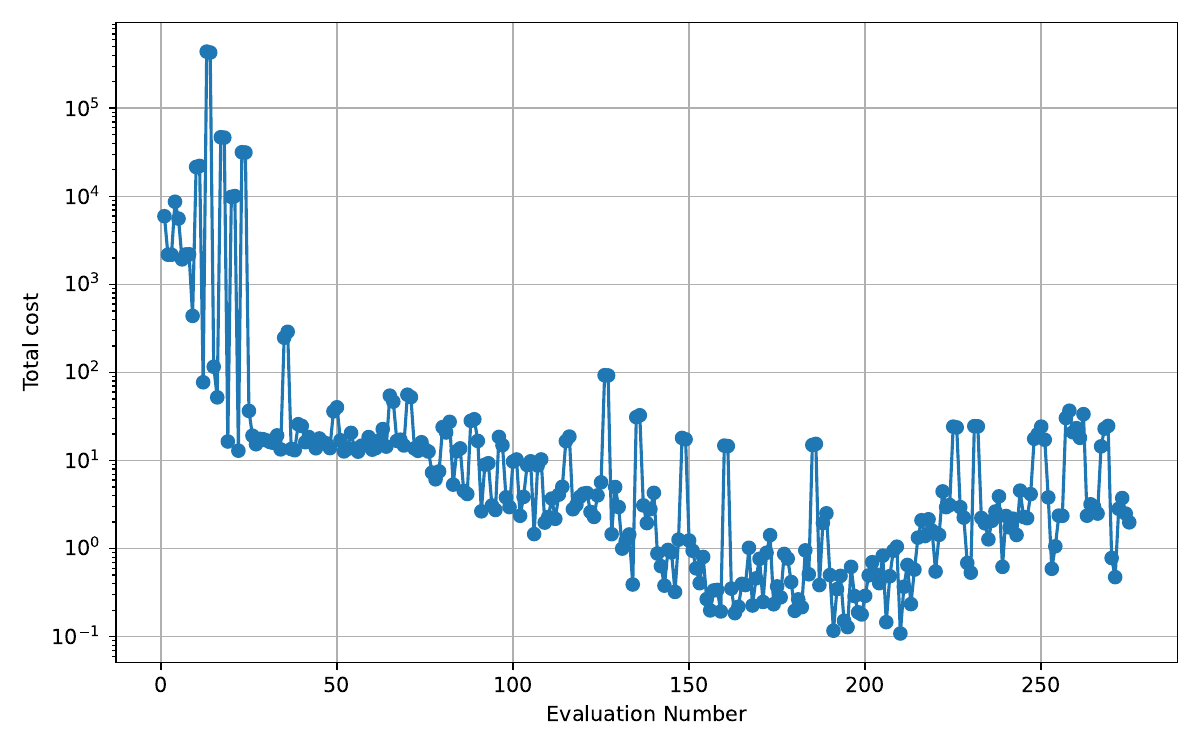}
\end{minipage}\hfill
\begin{minipage}[t]{0.333\linewidth}
  \centering
  \includegraphics[page=2,width=\linewidth]{RealHWMD.pdf}
\end{minipage}\hfill
\begin{minipage}[t]{0.333\linewidth}
  \centering
  \includegraphics[page=3,width=\linewidth]{RealHWMD.pdf}
\end{minipage}
\caption{Real-hardware H-DES results for the hypoelastic 1-D tensile test: convergence 
and comparison to the analytical reference.}
\label{fig:md_results}
\end{figure}

\paragraph{Hardware execution and results.}
We report results for the parameter set 
\(\epsilon_0=0.5\), \(\sigma_0=5\), \(b=10\), \(K=100\), \(n=4\), and \(g=12\).
We use \(n_1=15\) qubits for the local encoding of \(u(x)\) and \(n_2=4\) qubits for the 
global encoding of \(\sigma(x)\), with circuit depths 2 and 4, respectively. Each loss 
evaluation uses up to 20000 shots per function evaluation. The run was executed on IBM Marrakesh (Heron R2) on 
May~25,~2025. The convergence behavior and agreement with the analytical reference are 
shown in Fig.~\ref{fig:md_results}. The reference solutions here are $u(x)=\dfrac{16x^5}{45\sqrt{3}}-\dfrac{32x^4}{15\sqrt{3}}+\dfrac{128x^3}{25\sqrt{3}}-\Big(\dfrac{1}{20}+\dfrac{256\sqrt{3}}{125}\Big)x^2+\Big(\dfrac{3}{25}+\dfrac{768\sqrt{3}}{625}\Big)x$ and $\sigma(x)=-10x+12$.

The configuration for this run is given in Table \ref{tab:md_config} and the results for $u$ and $\sigma$ provided in Fig. \ref{fig:md_results}. The loss exhibits good convergence properties, decreasing by several orders of magnitude within a reasonable number of iterations, with no evidence of a barren plateau. The minimum value obtained is on the order of $10^{-1}$. This value does not correspond to the final value of the loss (at optimizer termination) due to errors arising from shot noise. The minimum loss value is sufficiently small to achieve high accuracy for the linear solution of 
$\sigma(x)$, but it is not low enough for a perfect matching between the analytical and numerical approximation of $u(x)$.

\begin{table}[htbp]
\centering
\begin{tabular}{|l|l|}
\hline
\textbf{Material deformation benchmark} & \textbf{Configuration} \\
\hline
Backend & IBM Marrakesh (Heron R2) \\
Ansatz & HEA  \\
Qubits & \(n_1=15\) (local), \(n_2=4\) (global) \\
Circuit depth & 2 (\(u\)), 4 (\(\sigma\)) \\
Observable encoding & Local + global \\
BC strategy & BC shift  \\
Optimizer & SLSQP (gradient-based) \\
Gradient evaluation & Parameter-shift rule \\
Shots per evaluation & Up to 20000 \\
Run date & May 25, 2025 \\
\hline
\end{tabular}
\caption{Configuration summary for the hypoelastic 1-D tensile test.}
\label{tab:md_config}
\end{table}

\subsection*{Inviscid Burgers' equation\\}
\vspace{-2mm}

\paragraph{Model and analytical solution.}
We consider the inviscid Burgers' equation
\begin{equation}\label{eq:burgers}
\frac{\partial u}{\partial t}+u\frac{\partial u}{\partial x}=0,
\qquad
u(x,0)=f(x),
\end{equation}
with linear initial conditions \(f(x)=ax+b\). The parameters \((a,b)\) control the 
magnitude of spatial and temporal gradients and therefore strongly influence the 
expressivity requirements of the variational ansatz and the sensitivity of the 
optimization to hardware noise. For this initial condition $f$, the analytical solution is $u(x,t)=\dfrac{ax+b}{at+1}$.

\paragraph{Optimization strategy under hardware noise.}

In practice, shot noise remains one of the dominant sources of uncertainty when 
executing variational quantum algorithms on hardware. Increasing the number 
of shots per expectation-value estimate improves accuracy but also significantly 
increases QPU execution time and cost. To balance this trade-off, we employ a multi-stage 
(N-stage) sampling strategy\cite{}, in which the optimization begins with a small number of 
shots to explore the parameter space and progressively increases the shot count as the 
optimizer approaches regions of lower loss. This strategy \cite{shots2020,shots2023} reduces the total number of function evaluations required to reach convergence while maintaining sufficient accuracy 
near the optimum.

In addition, the population-based nature of CMA-ES\cite{Hansen2023} makes it particularly well suited to 
this setting: the optimizer can tolerate noisy loss evaluations and does not rely on 
explicit gradient information, which becomes unreliable in the presence of hardware and 
shot noise. To further reduce variance in the loss estimates without increasing circuit 
depth, we employ a parallel stacking strategy in which multiple identical circuit blocks are 
executed in parallel on disjoint subsets of qubits, and their measurement outcomes are 
aggregated. Together, these techniques proved essential for achieving stable convergence 
of the inviscid Burgers' equation on real quantum hardware.

\begin{figure}[!ht]
\begin{center}
\begin{minipage}{0.48\textwidth}
\[
\Qcircuit @C=0.8em @R=0.8em {
& \lstick{q_0} & \gate{R_y(\theta_{1})} & \ctrl{1} & \qw      & \gate{R_x(\theta_{5})} & \ctrl{1}  & \qw       & \gate{R_y(\theta_{9})}  & \ctrl{1}  & \qw       & \multigate{3}{\mathcal{O}} \\
& \lstick{q_1} & \gate{R_y(\theta_{2})} & \targ    & \ctrl{1} & \gate{R_x(\theta_{6})} & \ctrl{-1} & \ctrl{1}  & \gate{R_y(\theta_{10})} & \ctrl{-1} & \ctrl{1}  & \ghost{\mathcal{O}}        \\
& \lstick{q_2} & \gate{R_y(\theta_{3})} & \ctrl{1} & \targ    & \gate{R_x(\theta_{7})} & \ctrl{1}  & \ctrl{-1} & \gate{R_y(\theta_{11})} & \ctrl{1}  & \ctrl{-1} & \ghost{\mathcal{O}}        \\
& \lstick{q_3} & \gate{R_y(\theta_{4})} & \targ    & \qw      & \gate{R_x(\theta_{8})} & \ctrl{-1} & \qw       & \gate{R_y(\theta_{12})} & \ctrl{-1} & \qw       & \ghost{\mathcal{O}}
}
\]
\end{minipage}
\qquad\qquad
\begin{minipage}{0.25\textwidth}
\centering
\[
\Qcircuit @C=1em @R=0.8em {
\lstick{q_0}  & \multigate{3}{\text{Structure 1}} & \qw \\
\lstick{q_1}  & \ghost{\text{Structure 1}} & \qw \\
\lstick{q_2}  & \ghost{\text{Structure 1}} & \qw \\
\lstick{q_3}  & \ghost{\text{Structure 1}} & \qw \\
\vdots \\
\lstick{q_{36}}  & \multigate{3}{\text{Structure 10}} & \qw \\
\lstick{q_{37}} & \ghost{\text{Structure 10}} & \qw \\
\lstick{q_{38}} & \ghost{\text{Structure 10}} & \qw \\
\lstick{q_{39}} & \ghost{\text{Structure 10}} & \qw \\
}
\]
\end{minipage}
\end{center}
\caption{\textbf{Left:} HEA structure pre-stacking two-qubit CNOT and CZ gates. The last block represents additional operations required for the estimation of expectation values of the observables. \textbf{Right:} The full circuit to obtain satisfactory solutions for the inviscid Burgers' equation with 40 qubits.}
\label{fig:side_by_side_structures}
\end{figure}
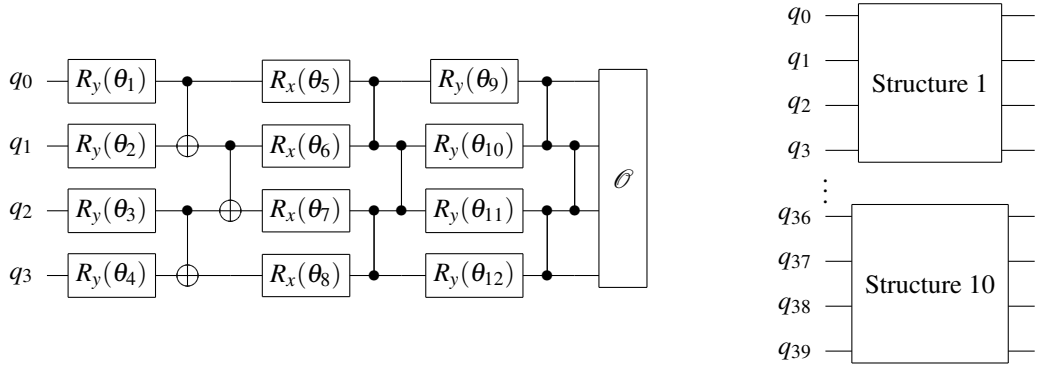

\paragraph{Ansatz selection — Case 1:} For \(a=0.5\),\, \(b=0.25\), we employ a variation of the HEA we used above with a global observable $\mathcal{O}$ (Eq. (\ref{eq:globalobs})). We set $n_1=1$ qubit for the $x$-variable, $n_2=2$ qubits for the $t$-variable with an extra qubit required for the global observable, combined with a stacked execution strategy 
to budget the total number of shots for the entire algorithm execution and reduce the runtime of the execution. The resulting 40-qubit stacked 
circuit is shown in Fig.~\ref{fig:side_by_side_structures}. This configuration was chosen 
based on empirical convergence speed and stability observed on hardware.

\paragraph{Results — Case 1: \(a=0.5\), \(b=0.25\).}
The run was executed on \texttt{ibm\_fez} (Heron~R2)  with 
collocation grids of 30 sample points in \(x\in[0,0.95]\) and 51 sample points in \(t\in [0,0.95]\). 
The shot schedule and attained minimum loss values are summarized in 
Table~\ref{tab:shot_noise}. 

\begin{table}[htbp]
\centering
\begin{tabular}{|c|c|c|c|c|}
\hline
\textbf{Stage} & 1 & 2 & 3 & 4 \\
\hline
Shots / eval & 500 & 2500 & 5000 & 10000 \\
\hline
\(\sigma_0\) & 0.5 & 0.25 & 0.1 & 0.05 \\
\hline
Avg. QPU time / eval [s] & 1 & 2 & 2 & 4 \\
\hline
Lowest loss & 0.05595382 & 0.006102567 & 0.002834661 & 0.001771369 \\
\hline
\end{tabular}
\caption{N-stage shot allocation and achieved loss minima for Burgers' equation 
(\(a=0.5\), \(b=0.25\)).}
\label{tab:shot_noise}
\end{table}

\begin{figure}[!ht]
  \centering
  \includegraphics[width=1.0\linewidth]{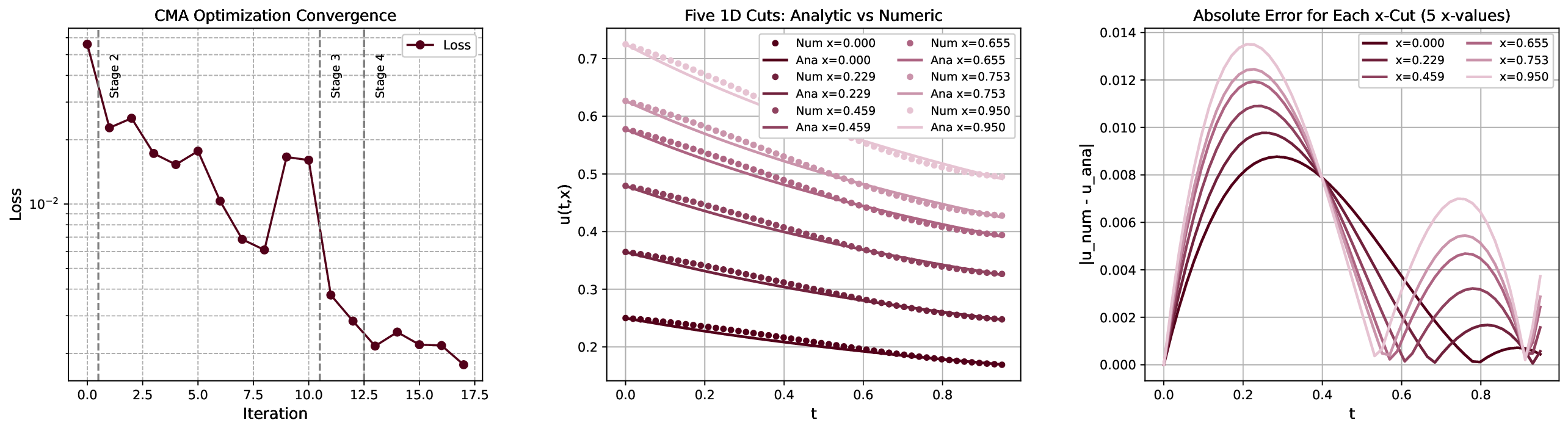}
  \caption{H-DES results for inviscid Burgers' equation on \texttt{ibm\_fez} (Heron~R2) 
  using 40 qubits and a 4-stage optimization strategy. \textbf{Left:} loss convergence 
  across stages. \textbf{Center:} reconstructed solution \(u(x,t)\). \textbf{Right:} 
  absolute error along selected \(x\)-cuts.}
  \label{fig:burgers_case1}
\end{figure}

Convergence and solution accuracy are shown in 
Fig.~\ref{fig:burgers_case1}. The results were obtained on Dec~5,~2025. The loss function converges to the desired accuracy after 17 total iterations across stages. The staged strategy improves convergence efficiency by progressively increasing the number of shots per stage (500, 1000, 5000, and 10 000) thereby balancing QPU time and accuracy, with early stages characterized by low QPU time and coarse accuracy and later stages by higher QPU time and improved solution quality. The steep decrease of the loss function, with no indication of a plateau, suggests that further improvements in solution resolution are achievable by increasing the number of iterations and shots.

\begin{figure}[!ht]
  \begin{center}
  \[
\hspace{2.3cm}\Qcircuit @C=0.5em @R=0.5em {
& \lstick{q_0} & \gate{R_x(\theta_1)}  & \ctrl{1} & \qw      & \qw      & \qw      & \gate{R_x(\theta_6)}    & \ctrl{1} & \qw      & \qw      & \qw      & \gate{R_x(\theta_{11})} & \ctrl{1} & \qw      & \qw      & \qw     & \gate{R_x(\theta_{16})} & \ctrl{1} & \qw      & \qw      & \qw      & \multigate{4}{\mathcal{O}} \\
& \lstick{q_1} & \gate{R_x(\theta_2)}  & \ctrl{0} & \ctrl{1} & \qw      & \qw      & \gate{R_x(\theta_7)}    & \ctrl{0} & \ctrl{1} & \qw      & \qw      & \gate{R_x(\theta_{12})} & \ctrl{0} & \ctrl{1} & \qw      & \qw     & \gate{R_x(\theta_{17})} & \ctrl{0} & \ctrl{1} & \qw      & \qw      & \ghost{\mathcal{O}}        \\
& \lstick{q_2} & \gate{R_x(\theta_3)}  & \qw      & \ctrl{0} & \ctrl{1} & \qw      & \gate{R_x(\theta_8)}    & \qw      & \ctrl{0}  & \ctrl{1} & \qw      & \gate{R_x(\theta_{13})} & \qw      & \ctrl{0} & \ctrl{1} &\qw      & \gate{R_x(\theta_{18})} & \qw      & \ctrl{0} & \ctrl{1} & \qw      & \ghost{\mathcal{O}}        \\
& \lstick{q_3} & \gate{R_x(\theta_4)}  & \qw      & \qw      & \ctrl{0} & \ctrl{1} & \gate{R_x(\theta_9)}    & \qw      & \qw      & \ctrl{0} & \ctrl{1} & \gate{R_x(\theta_{14})} & \qw      & \qw      & \ctrl{0} &\ctrl{1} & \gate{R_x(\theta_{19})} & \qw      & \qw      & \ctrl{0} & \ctrl{1} & \ghost{\mathcal{O}}        \\ 
& \lstick{q_4} & \gate{R_x(\theta_5)}  & \qw      & \qw      & \qw      & \ctrl{0} & \gate{R_x(\theta_{10})} & \qw      & \qw      & \qw      & \ctrl{0} & \gate{R_x(\theta_{15})} & \qw      & \qw      & \qw      &\ctrl{0} & \gate{R_x(\theta_{20})} & \qw      & \qw      & \qw      & \ctrl{0} & \ghost{\mathcal{O}}
}
\]
\end{center} 
  \caption{Hardware-efficient \(R_x\)+\(CZ\) ansatz used for Burgers' equation (\(a=1.0\), \(b=1.0\)).}
  \label{fig:burgers_ansatz_rxcz}
\end{figure}
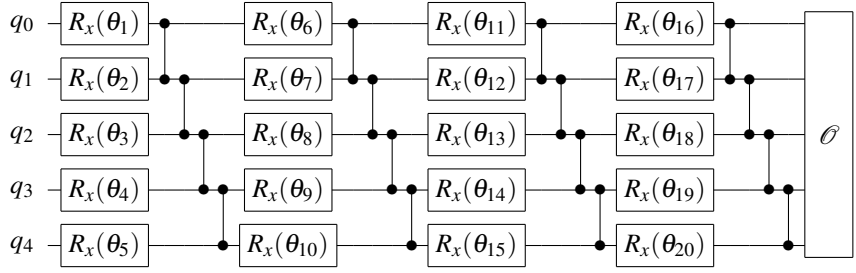

\paragraph{Ansatz selection — Case 2:} For the choice of parameters $a = 1$ and $b = 1$, convergence depends more strongly on the expressivity of the ansatz and 
on the initialization of circuit parameters. We use $n_1=1$ qubit for the $x$-variable and $n_2=3$ qubits for the $t$-variable and a hardware-efficient architecture 
based on parameterized \(R_x\) rotations and \(CZ\) entanglers (cascade), shown in 
Fig.~\ref{fig:burgers_ansatz_rxcz} followed by the stacking strategy. The purpose here is to reduce gate noise which increases with the transpilation process when we don't use native gates. Replacing \(R_y\) and $CNOT$ by \(R_x\) and $CZ$ allows us to avoid the use of error mitigation techniques for the chosen hardware. This alternative choice of  ansatz is possible because the expressivity remains sufficient for this problem. Circuit parameters are initialized using backend-aware heuristics derived from prior runs.

\begin{figure}[H]
  \centering
  \includegraphics[width=1.0\linewidth]{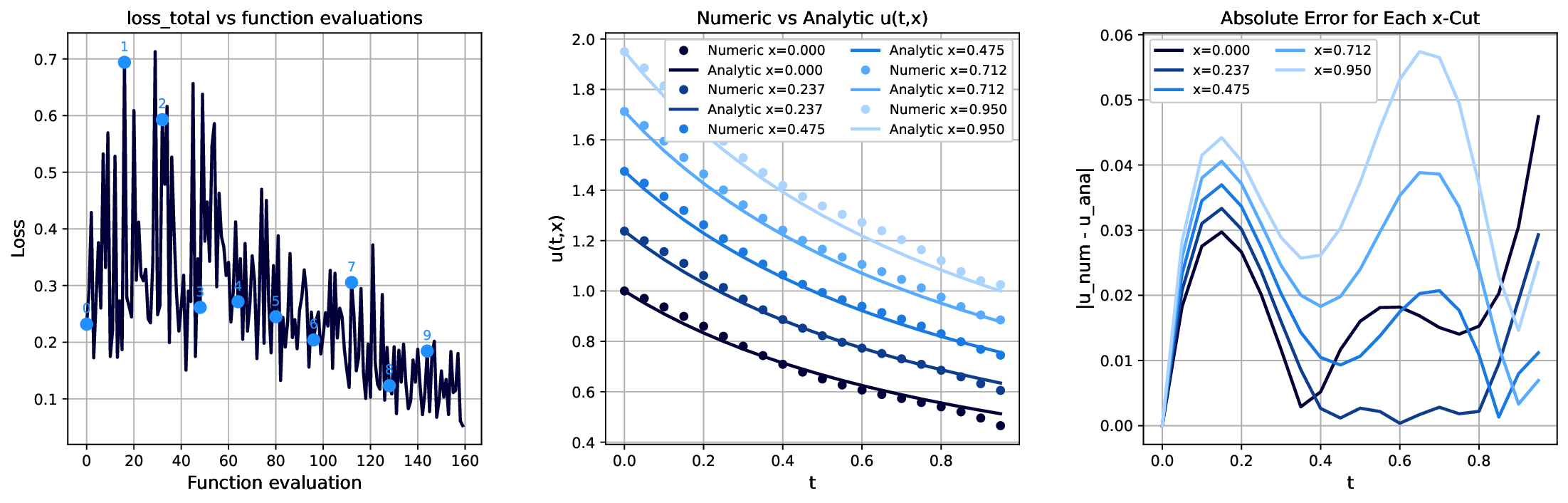}
  \caption{H-DES results for inviscid Burgers' equation on \texttt{ibm\_fez} (Heron~R2) 
  using 50 qubits and 10000 shots per function evaluation. \textbf{Left:} loss 
  convergence. \textbf{Center:} reconstructed solution \(u(x,t)\). \textbf{Right:} 
  absolute error along selected \(x\)-cuts.}
  \label{fig:burgers_case2}
\end{figure}

\paragraph{Results — Case 2: \(a=1.0\), \(b=1.0\).}
The run uses 50 qubits, four ansatz layers, and 10000 shots per function evaluation.  The execution was carried out on 
\texttt{ibm\_fez} (Heron~R2) and reaches a best loss value of 
\(5.27\times 10^{-2}\) after 10 iterations (160 evaluations). The results were obtained 
on Dec~14,~2025 and are shown in Fig.~\ref{fig:burgers_case2}. 

Using a single-stage strategy with 10000 shots initialized from a high-quality seed, the loss function decreases steadily without reaching a plateau. The optimization rapidly converges within a limited number of iterations, achieving the target accuracy, with the final objective value coinciding with the best-ever value observed during the run. This behavior highlights that, when starting from a good initial seed, a high-shot strategy can yield fast and stable convergence without the need for multi-stage shot scheduling.

\begin{table}[htbp]
\centering
\begin{tabular}{|l|l|}
\hline
\textbf{Inviscid Burgers' benchmark} & \textbf{Configuration} \\
\hline
Backend & \texttt{ibm\_fez} (Heron R2) \\
Ansatz & HEA (stacked) / \(R_x+CZ\) cascade \\
Qubits & 40 (case 1), 50 (case 2) \\
Circuit depth & 3--4 layers \\
Observable encoding & Chebyshev spectral \\
Boundary conditions & Shift-based \\
Optimizer & CMA-ES (derivative-free) \\
Shot strategy & N-stage sampling \\
Shots per evaluation & 500--10000 \\
Execution mode & \texttt{Session} \\
Run dates & Dec 5 \& Dec 14, 2025 \\
\hline
\end{tabular}
\caption{Configuration summary for the inviscid Burgers' equation benchmarks.}
\label{tab:burgers_config}
\end{table}

\begin{figure}[!ht]
\centering
\includegraphics[width=1.0\linewidth]{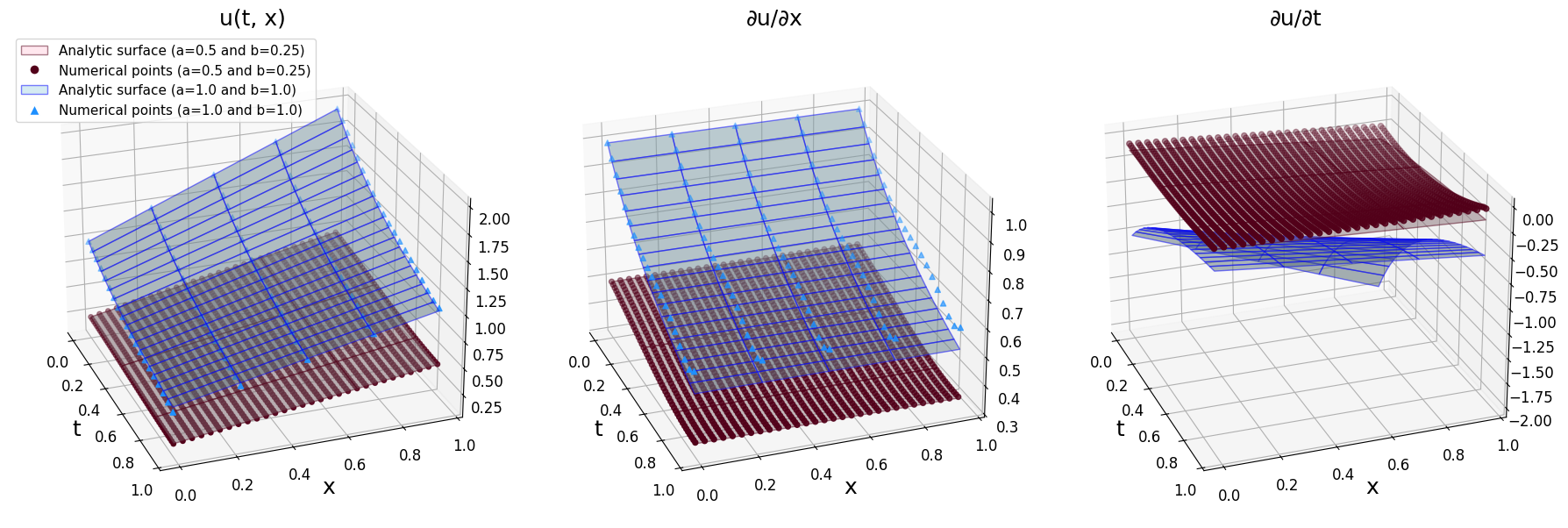} 
\caption{Comparison of converged inviscid Burgers' solutions for two initial conditions: 
\(a=0.5,\,b=0.25\) (dark-red) and \(a=1.0,\,b=1.0\) (blue). \textbf{Left:} recovered solutions and analytic 
reference. \textbf{Center:} postprocessed \(\partial_x u\). \textbf{Right:} postprocessed 
\(\partial_t u\) (note the limited temporal resolution for the \(a=1.0,\,b=1.0\) run).}
\label{fig:burgers_compare}
\end{figure}

\paragraph{Effect of initial-condition parameters.}
Figure~\ref{fig:burgers_compare} compares the recovered solution surfaces and 
postprocessed derivatives for the two converged cases. As \((a,b)\) increase, the 
solution exhibits larger derivatives in both space and time, increasing the number of 
degrees of freedom required to approximate \(u(x,t)\) and its derivatives across the 
domain. This translates into the need for more qubits (higher spectral order), more 
expressive ansatz architectures, and more careful optimization and initialization 
strategies on noisy quantum hardware. 

\section*{Discussion}

The results presented in this work demonstrate that H-DES is, by design, well suited to 
the constraints of present-day Noisy Intermediate-Scale Quantum (NISQ) hardware. Across 
both benchmarks, the algorithm exhibits a high degree of robustness to hardware noise, 
achieving convergence without relying heavily on explicit error mitigation techniques. 
Instead, hardware noise effects are  absorbed  through the variational loss 
minimization process itself. This 
property significantly accelerates the overall execution workflow and reduces the 
overhead typically associated with error-mitigation pipelines, while still enabling 
accurate solutions of nontrivial differential equations on real quantum processors.

Noise resilience in variational quantum algorithms is not achieved through a single universal 
strategy, but rather through a combination of problem-adapted choices. In the case of 
the hypoelastic ODE system, the smoothness of the solution and the  low 
dimensionality of the problem allowed for shallow circuits and gradient-based 
optimization using the parameter-shift rule. In this regime, hardware noise remained 
sufficiently small that convergence could be achieved without mitigation and 
standard shot budgets. On the other hand, the inviscid Burgers' equation posed a more 
challenging optimization landscape, due to higher-dimensional 
function representations. In this case, 
derivative-free optimization (CMA-ES), multi-stage shot allocation, and circuit stacking 
proved essential to stabilize and accelerate the training process and achieve convergence on hardware.

These observations highlight an important distinction between noise sources in quantum 
algorithms. While gate and readout errors can often be mitigated by improvements in 
hardware quality or modest resilience settings, shot noise remains a fundamental 
bottleneck for both NISQ and future fault-tolerant quantum computers. The results of this 
work indicate that effective management of shot noise requires a global approach, 
involving the joint design of the ansatz, the observable encoding, the loss function, 
the measurement strategy, and the classical optimizer. No single component alone is 
sufficient; rather, robustness emerges from their combined interaction within the 
hybrid quantum--classical loop.

Although the examples presented here did not require  error mitigation, we 
emphasize that this may not hold universally. In particular, problems requiring deeper 
circuits, higher spectral resolution, or more complex observables may benefit from 
additional mitigation techniques\cite{tousi2025quantum}. Nevertheless, the ability of H-DES to operate 
efficiently on current NISQ devices, while maintaining flexibility in its configuration, 
suggests promising scaling behavior as hardware improves. More broadly, these results 
support the view that variational quantum differential equation solvers should be treated as adaptable toolboxes rather than fixed algorithms, with optimal performance emerging from problem-specific and hardware-aware design choices.

It is instructive to place the present results in the context of other recent attempts 
to execute variational or hybrid quantum differential equation solvers on real quantum 
hardware. Variational quantum linear solvers \cite{BravoPrieto2023}, nonlinear variational 
frameworks \cite{Lubasch2020}, and hybrid quantum–classical schemes for fluid dynamics 
problems \cite{Song2024NavierStokes} have demonstrated the conceptual feasibility of such 
approaches, but have also highlighted practical challenges when deployed on NISQ 
processors. In particular, several studies report sensitivity of the optimization 
process to shot noise, circuit depth, and optimizer choice, often resulting in slow or 
unstable convergence on current quantum devices\cite{ali2023performance}.

The results presented here suggest that these challenges are not inherent to the 
variational paradigm itself, but rather depend critically on how convergence is handled within 
the hybrid loop. By treating the ansatz, observable encoding, loss construction, shot 
allocation, and optimizer choice as jointly tunable components, H-DES enables convergence 
in regimes where more rigid formulations struggle. This is consistent with recent 
observations in quantum circuit learning and physics-informed approaches 
\cite{Schillo2024,Panichi2025}, which emphasize the importance of adaptive design choices 
for successful execution on NISQ hardware.

Importantly, our findings do not contradict earlier reports of convergence difficulties; 
instead, they provide evidence that such difficulties can be mitigated through 
problem-aware and hardware-aware algorithmic design. In this sense, the present work 
demonstrates a complementary pathway toward practical quantum differential equation 
solvers, focused on robustness and flexibility rather than asymptotic guarantees.

In future work, one intends to tackle more complex differential problems and test alternative choices of ansatz, observables, mitigation and optimization strategies, and to run our experiments on other various quantum devices. We believe that increasing the variety of real problems solved on quantum backends is a valuable strategy to prove  near term quantum utility.


\section*{Acknowledgments}

 We acknowledge the use of IBM Quantum services for this work. The views expressed are those of the authors and do not reflect the official policy or position of IBM or the IBM Quantum team. The authors would like to thank the developers of the open-source framework Qiskit as well as the developers of the library MPQP by ColibriTD that was used to write a primilary version of the code. The authors acknowledge the use of large language model–based tools to assist with 
language editing. All scientific content, interpretations, 
and conclusions are the sole responsibility of the authors.

\section*{Author contributions statement}

K.B., Y.M., R.R., R.K. and A.B. developed the main code, K.B. and Y.M conducted the experiments on the IBM Quantum Platform. All authors analyzed the results, reviewed and contributed to the manuscript. 
\section*{Data availability}
All experimental data can be provided under request to the corresponding author. The algorithms presented in this work have been packaged into a Qiskit function \href{https://quantum.cloud.ibm.com/docs/en/guides/colibritd-pde}{QUICK PDE} available through IBM Quantum plans.

\section*{Funding}
Open Access funding covered by ColibriTD. The authors did not recevie any external funding for this works. Additionally, the authors declare that they have no conflicts of interest to disclose.

\bibliography{sample}
\end{document}